\documentclass[lettersize,journal]{IEEEtran}
\usepackage{amsmath,amsfonts}
\usepackage{algorithmic}
\usepackage{algorithm}
\usepackage{array}
\usepackage[caption=false,font=normalsize,labelfont=sf,textfont=sf]{subfig}
\usepackage{textcomp}
\usepackage{stfloats}
\usepackage{url}
\usepackage{verbatim}
\usepackage[dvipdfmx]{graphicx}
\usepackage{cite}
\usepackage{siunitx} 
\usepackage{bm} 
\usepackage{multirow} 
\usepackage{color}

\begin{document}

\title{Fabrication of a 64-Pixel TES Microcalorimeter Array with Iron Absorbers Uniquely Designed \\for 14.4-keV Solar Axion Search}

\author{Yuta Yagi, Tasuku Hayashi, Keita Tanaka, Rikuta Miyagawa, Ryo Ota, Noriko Y. Yamasaki, Kazuhisa Mitsuda, Nao Yoshida, Mikiko Saito, and Takayuki Homma
\thanks{Manuscript received 14 November 2022; revised 27 December 2022; accepted 2 January 2023. Date of publication 8 March 2023.}

\thanks{Most of the fabrication in this work was performed in the nano-electronics clean room at the Institute of Space and Astro- nautical Institute, Japan Aerospace Exploration Agency. In addition, iron electroplating was performed at the Nanotechnology Research Center at Waseda University. The niobium lead was deposited by sputtering a clean room at the Advanced Technology Center, National Astronomical Observatory of Japan. The authors would like to thank the technical staff of these facilities. Furthermore, the Japan Society for the Promotion of Science (JSPS) KAKENHI Grant Number 20H05857 and 20K14548 financially supported this work.}

\thanks{Y. Yagi  (e-mail: yagi@ac.jaxa.jp), K. Tanaka, R. Miyagawa, and R. Ota are with the Department of Space Astronomy and Astrophysics, the Institute of Space and Astronautical Science (ISAS), Japan Aerospace Exploration Agency (JAXA), 3-1-1 Yoshinodai, Chuo-ku, Sagamihara-shi, Kanagawa 252-5210, Japan, and also with the Department of Physics, School of Science, the University of Tokyo, 7-3-1 Hongo, Bunkyo-ku, Tokyo 113-0033, Japan.}
\thanks{T. Hayashi is with the International Center for Quantum-field Measurement Systems for Studies of the Universe and Particles (QUP), KEK, 1-1 Oho, Tsukuba-shi, Ibaraki 305-0801, Japan, and also with the National Astronomical Observatory of Japan (NAOJ), 2-21-1 Osawa, Mitaka-shi, Tokyo 181-8588, Japan.}
\thanks{N. Y. Yamasaki is with the Department of Space Astronomy and Astrophysics, ISAS, JAXA, with the Department of Physics, School of Science, The University of Tokyo, and also with QUP, KEK.}
\thanks{K. Mitsuda is with NAOJ, with QUP, KEK, and also with ISAS, JAXA.}
\thanks{N. Yoshida is with the Department of Applied Chemistry,  School of Advanced Science and Engineering, Waseda University, Okubo 3-4-1, Shinjuku-ku, Tokyo 169-8555, Japan.}
\thanks{M. Saito is with the Research Organization for Nano \& Life Innovation, Waseda University, 513 Waseda-tsurumaki-cho, Shinjuku-ku, Tokyo 162-0041, Japan.}
\thanks{T. Homma is with the Department of Applied Chemistry,  School of Advanced Science and Engineering, Waseda University, and also with the Research Organization for Nano \& Life Innovation, Waseda University.}
}

\markboth{IEEE TRANSACTIONS ON APPLIED SUPERCONDUCTIVITY,~Vol.33, No.~5, AUGUST~2023}%
{Shell \MakeLowercase{\textit{et al.}}: A Sample Article Using IEEEtran.cls for IEEE Journals}


\maketitle

\begin{abstract}
If a hypothetical elementary particle called an axion exists, to solve the strong CP problem, a $^{57}$Fe nucleus in the solar core could emit a 14.4-keV monochromatic axion through the M1 transition. If such axions are once more transformed into photons by a $^{57}$Fe absorber, a transition edge sensor (TES) X-ray microcalorimeter should be able to detect them efficiently. We have designed and fabricated a dedicated 64-pixel TES array with iron absorbers for the solar axion search. In order to decrease the effect of iron magnetization on spectroscopic performance, the iron absorber is placed next to the TES while maintaining a certain distance. A gold thermal transfer strap connects them. We have accomplished 
the electroplating of gold straps with high thermal conductivity. The residual resistivity ratio (RRR) was over 23, more than eight times higher than a previous evaporated strap. In addition, we successfully electroplated pure-iron films of more than a few micrometers in thickness for absorbers and a fabricated 64-pixel TES calorimeter structure.
\end{abstract}


\newpage
\section{Introduction}
\IEEEpubidadjcol
\IEEEPARstart{A}{n} axion is a hypothetical elementary particle proposed to solve the strong CP problem in particle physics \cite{PecceiandQuinn77, WeinbergandWilczek78} and is one of the cold dark-matter candidates in the Universe in astrophysics. In the solar core, a $^{57}$Fe nucleus interacting with blackbody photon with $\SI{1.3}{keV}$\cite{Turck+93} is excited to the 14.4-keV first level through the M1 transition. However, it deexcites soon and emits a 14.4-keV monochromatic axion. At the reconversion, the $^{57}$Fe emits a 14.4-keV $\gamma$-ray, conversion electron, or a lower energy X-ray with a certain probability. There are several trials to detect solar axions on the ground. We focus on 14.4-keV axions produced by the $^{57}$Fe M1 transition. High energy-resolution transition edge sensor (TES) microcalorimeter with $^{57}$Fe absorbers can detect not only 14.4-keV $\gamma$-rays but also others as thermal energy. Therefore, the TES microcalorimeter is expected to have a much higher sensitivity \cite{Yagi+23} than a semiconductor sensor which can detect only the 14.4-keV $\gamma$-rays escaped from $^{57}$Fe \cite{Krcmar+98, Derbin+07, Namba07, Derbin+09, Derbin+11}. The best upper limit using $^{57}$Fe on axion mass of $m_{\rm a} < \SI{145}{eV}$ \cite{Derbin+11} is obtained at a 95\% confidence limit in the KSVZ hadronic axion model\cite{K79SVZ80}. On the other hand, 
another experiment
using a proportional gas chamber to detect 9.4-keV photons from $^{83}$Kr gas through the M1 transition bounded on the hadronic axion mass of $m_{\rm a} < \SI{12.7}{eV}$ (95\%  C.L.) \cite{Gavrilyuk15, Gavrilyuk18}.

We proposed that the TES microcalorimeter array with $^{57}$Fe absorbers could be a solar axion detector\cite{Konno+20, Yagi+23}. The iron absorber is set next to the TES with keeping a certain distance, and the TES and iron absorber are connected by a gold thermal transfer strap. This unique structure was chosen to avoid ferromagnetic iron's magnetization effect on the TES spectroscopic performance \cite{Ishisaki+08}. Konno et al. 2020 reported the degradation of transition sharpness using actual devices only with a $\num{60}$-$\si{\micro m}$ distance. 

This paper describes our development status of TES microcalorimeters with iron absorbers. We designed a 64-pixel TES microcalorimeter array with 
iron
absorbers and fabricated sensors with different distances between the TES and iron absorber from $\num{0}$ to $\SI{200}{\micro m}$ to evaluate the effect in detail. The high thermal conductivities of absorbers and thermal straps are required to achieve high energy resolution in the structure. We set conditions of the gold deposition for thermal straps by electroplating to achieve high thermal conductivity compared to 
the
former evaporated 
strap
\cite{Yagi+23}. In addition, iron-absorber fabrication by 
the
electroplating method is under study because it is an unusual process using pure iron on sub-mm scale patterns. Resistance values in this paper were measured using the LR-700 AC resistance bridge of Linear Research Inc.

\section{TES film by vacuum evaporation}

We introduced a Ti$/$Au vacuum evaporator (SEED Lab., Corporation) in 2021 to deposit TES at ISAS/JAXA. This equipment can deposit three 3-inch substrates in the same batch and automatically control the Ti$/$Au film 
thicknesses
and the deposition time interval between Ti and Au. The deposition  rates of Ti and Au films on the substrate were set to approximately 0.15 nm/s and 0.10 nm/s, respectively. In addition, the time from Ti deposition to the start of Au deposition was set to 55 s for ``SEED211026b" and ``SEED211109b" and 65 s for all other cases. The substrate ID numbers are the deposition date (yymmdd).

\begin{figure}[!htbp]
\centering
\includegraphics[trim=0mm 0mm 0mm 0mm, width=0.8\linewidth, keepaspectratio]{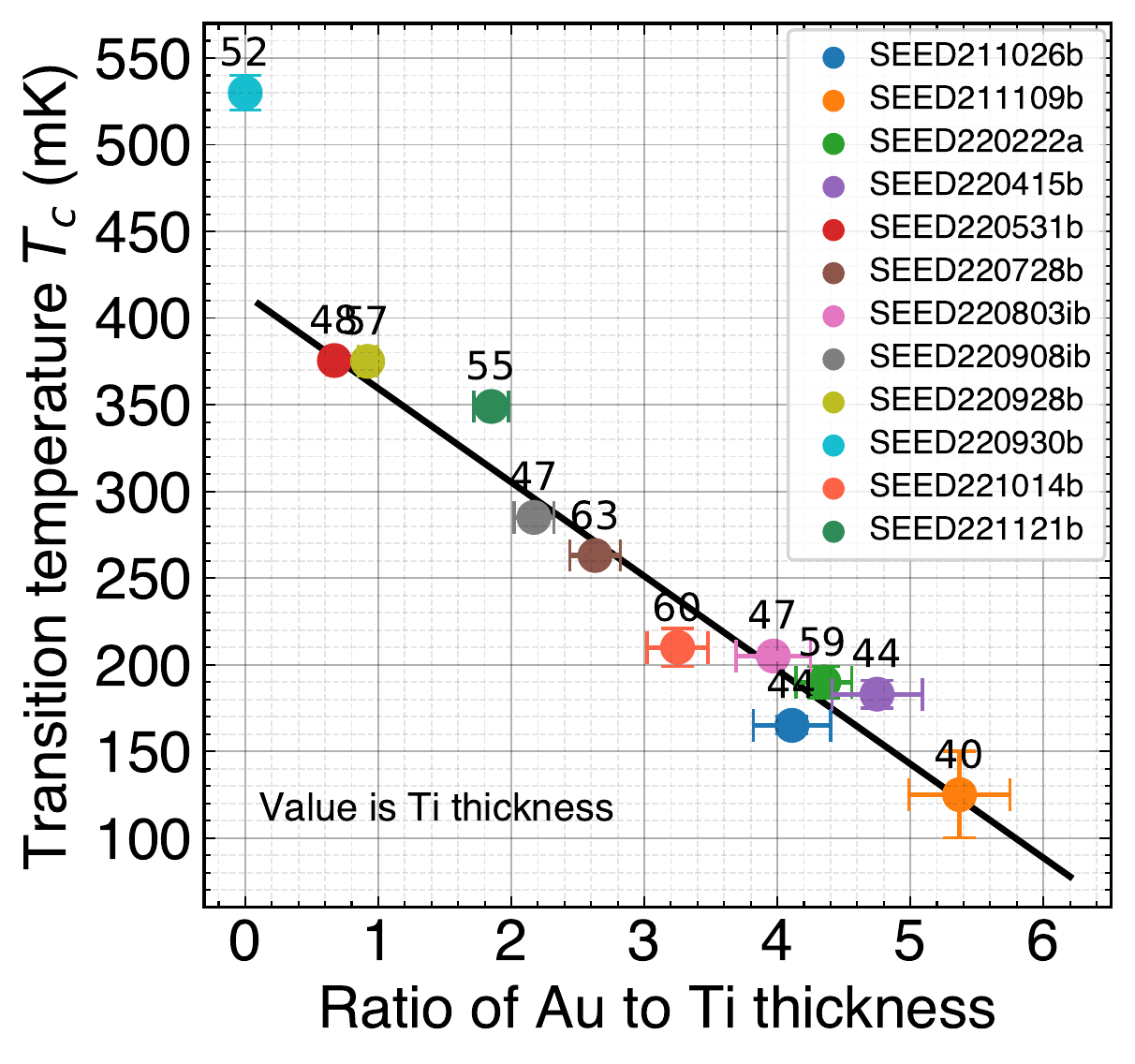}
\caption{Relation between the ratio of Au to Ti thickness and transition temperature $T_{\rm c}$. The value written above the plot point 
is
the thickness of the Ti film. The solid black line is a fitted linear line 
except for ``SEED220930b," which is titanium alone.}
\label{fig:TES_ratio_vs_Tc}
\end{figure}

In the beginning, the transition characteristics were checked. The TES was made $\SI{100}{\micro m}$ wide and $\SI{1000}{\micro m}$ long to facilitate the resistance measurement, measured using the four-terminal method. A current of $\SI{32}{\micro A}$ was applied to measure the relation between the electric resistance of the TES and the temperature (RT curve). The relation between the ratio of Au to Ti thickness and the transition temperature $T_{\rm c}$ of the deposited TES is shown in fig. \ref{fig:TES_ratio_vs_Tc}. The transition temperature of Ti alone ($\SI{52}{nm}$) was about 
$\num{520}$--$\SI{540}{mK}$.
The film thickness was the average of the thicknesses measured across the entire substrate. The variation of Ti and Au film thicknesses on the substrate was confirmed to be within $\pm$5\% of the averaged thickness, respectively, from which the error in the thickness ratio was calculated. Among the results of several elements for which the transition temperature $T_{\rm c}$ was measured, the central value was plotted as a representative value. The $T_{\rm c}$ error was determined so that all obtained $T_{\rm c}$ were included. However, the ruthenium oxide thermometer (Lake Shore Cryotronics Inc.) was not calibrated to measure the temperature. Therefore, the absolute temperature is inaccurate. The relation between the thickness ratio and $T_{\rm c}$ is linear, allowing the film thickness to be determined from the operating temperature when fabricating the calorimeter.

\section{Electroplating of Gold for Thermal Straps}
In Yagi et al. 2023 \cite{Yagi+23}, we had evaporated the gold thermal transfer strap by the electron beam physical vapor deposition. However, the electroplating generally makes 10 times higher thermal conductivity at 4K compared to the vapor deposition. The residual resistance ratio (RRR) of 300K to 4K, 
an
electrical and thermal conductivity
indicator,
is around 30 \cite{Nagayoshi+20}. Thermal simulations performed by Mori et al. 2022 \cite{Mori+22} reported that a 10-fold improvement in the thermal conductivity of the gold strap would reduce the position dependence in an iron absorber. Therefore, we introduced a plating environment at ISAS/JAXA and referred to Nagayoshi et al. 2020 \cite{Nagayoshi+20} to develop the conditions. 

\begin{table}[!htbp]
\caption{Conditions of gold electroplating, measured thicknesses, and measured resistances}
\centering
\scalebox{0.9}[0.9]{
\begin{tabular}{cccccc}
\hline
ID                                                                                       & E038                                                          & E039                                                           & E040                                                          & E042-1/-2                                                                              & E043-1/-2                                                                            \\ \hline
\vspace{+1mm}
\begin{tabular}[c]{@{}c@{}}Current \\ ($\si{mA}$)\end{tabular}                           & 0.140                                                         & 0.140                                                          & 0.070                                                         & 0.070                                                                                  & 0.070                                                                                \\
\vspace{+1mm}
\begin{tabular}[c]{@{}c@{}}Plated area \\ ($\si{cm^2}$)\end{tabular}                     & 0.112                                                         & 0.112                                                          & 0.112                                                         & 0.112                                                                                  & 0.112                                                                                \\
\vspace{+1mm}
\begin{tabular}[c]{@{}c@{}}Current \\ density \\ ($\si{mA/cm^2}$)\end{tabular}           & 1.25                                                          & 1.25                                                           & 0.63                                                          & 0.63                                                                                   & 0.63                                                                                 \\
\vspace{+1mm}
Time ($\si{min}$)                                                                        & 60                                                            & 35                                                             & 70                                                            & 70                                                                                     & 20                                                                                   \\
\vspace{+1mm}
\begin{tabular}[c]{@{}c@{}}Measured \\ thickness \\ ($\si{\micro m}$)\end{tabular}       & 3.64                                                          & 2.51                                                           & 2.40                                                          & \begin{tabular}[c]{@{}c@{}}2.06\\ /2.07\end{tabular}                                   & \begin{tabular}[c]{@{}c@{}}0.73\\  /0.74\end{tabular}                                \\
\vspace{+1mm}
\begin{tabular}[c]{@{}c@{}}Resistance \\ at room temp. \\ ($\si{m\Omega}$)\end{tabular}  & 61.5413                                                       & 103.702                                                        & 90.9307                                                       & \begin{tabular}[c]{@{}c@{}}98.8475\\ /97.4863\end{tabular}                             & \begin{tabular}[c]{@{}c@{}}233.330\\ /227.871\end{tabular}                           \\
\vspace{+1mm}
\begin{tabular}[c]{@{}c@{}}Resistance \\ at $\SI{4}{K}$ \\ ($\si{m\Omega}$)\end{tabular} & 4.41484                                                       & 12.0894                                                        & 3.64226                                                       & \begin{tabular}[c]{@{}c@{}}4.27643\\ /4.11929\end{tabular}                             & \begin{tabular}[c]{@{}c@{}}124.064\\ /117.779\end{tabular}                           \\
RRR                                                                                      & \begin{tabular}[c]{@{}c@{}}13.9397\\ $\pm$0.0139\end{tabular} & \begin{tabular}[c]{@{}c@{}}8.57795\\ $\pm$0.00410\end{tabular} & \begin{tabular}[c]{@{}c@{}}24.9655\\ $\pm$0.0274\end{tabular} & \begin{tabular}[c]{@{}c@{}}23.1145\\ $\pm$0.1015\\ /23.6658\\ $\pm$0.0955\end{tabular} & \begin{tabular}[c]{@{}c@{}}1.8807\\ $\pm$0.0794\\ /1.9347\\ $\pm$0.0992\end{tabular} \\ \hline
\end{tabular}
}
\label{tab:Au_ep_1}
\end{table}

\begin{figure}[!htbp]
\centering
\includegraphics[trim=0mm 0mm 0mm 0mm, width=0.68\linewidth, keepaspectratio]{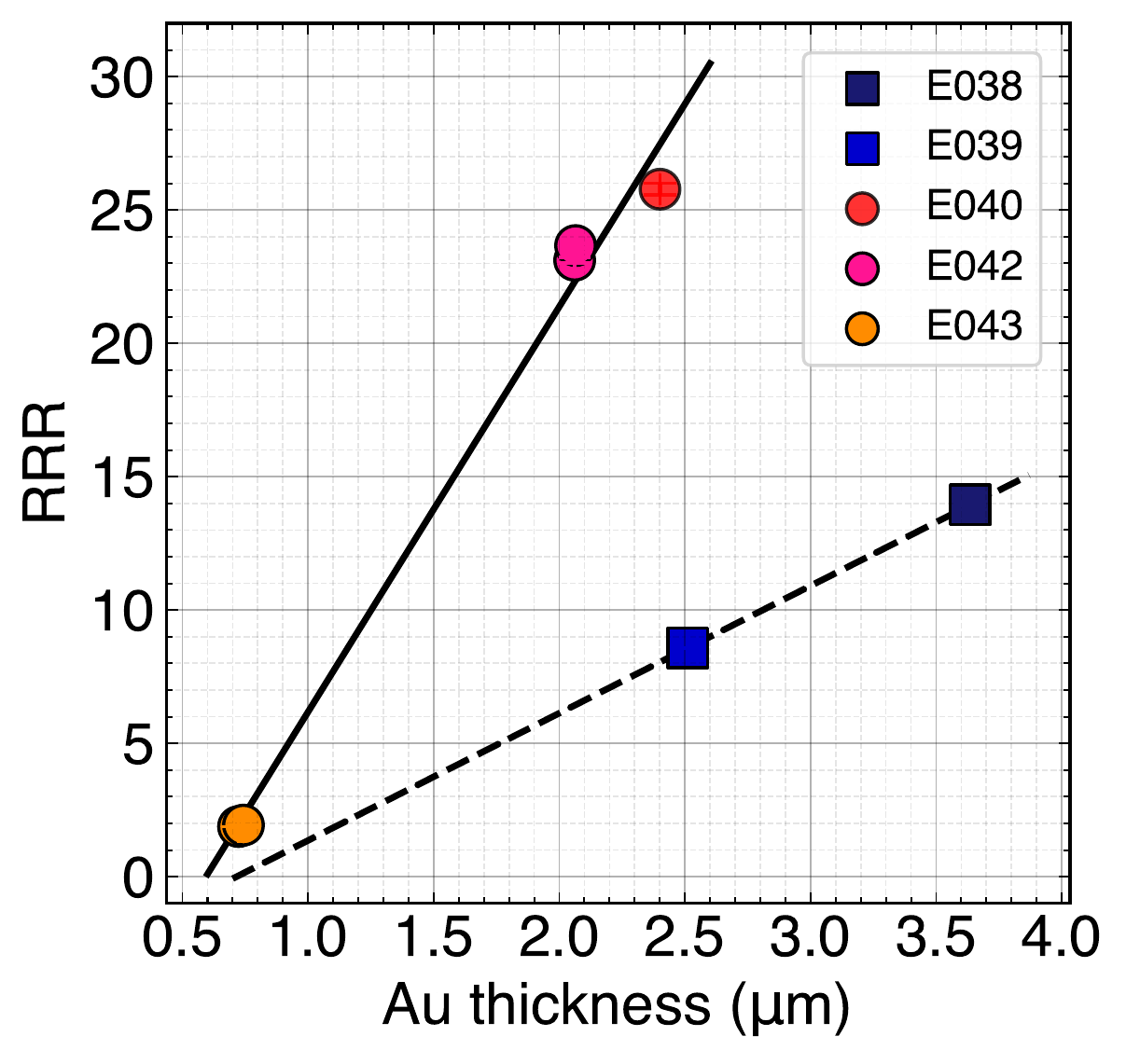}
\caption{Relation between film thickness and RRR at each current density. The squares and circles show groups deposited with $\SI{1.25}{mA/cm^2}$ and $\SI{0.63}{mA/cm^2}$, respectively. The solid and dashed lines show RRR-thickness relations for each group.}
\label{fig:Au_RRR_vs_thickness}
\end{figure}

``Techni Gold 25 ES RTU” (Technic Inc.) was used as the solution, and the film was deposited on a $\SI{2}{cm}$ square substrate to set the conditions. The seed layer was an evaporated Ti$/$Au film. The plating bath was heated to keep the solution temperature at $\num{40}$--$\num{45}$$^\circ$C, and the film was deposited while stirring at $\SI{360}{rpm}$. The deposition conditions, measured film thicknesses, and resistances are shown in tab.\ref{tab:Au_ep_1}. 
The stylus profilometer Dektak 6M (Veeco Instruments Inc.) measured the thicknesses.
The size
measured the resistance
of the gold film was $\SI{100}{\micro m}$ wide and $\SI{2000}{\micro m}$ long. In the RRR calculation, the resistance of the seed layer was negligible because the thickness
and resistance
of the seed layer
were enough
lower
and larger
than
those
of plated gold. The relation between film thickness and RRR is shown in fig. \ref{fig:Au_RRR_vs_thickness}. The trend varies with current density. A film with the required thickness of several micrometers and RRR$>$23 could be deposited by applying a constant current to achieve $\SI{0.63}{mA/cm^2}$. The plated films had eight times higher RRR than a previous evaporated gold film \cite{Yagi+23} and are expected to reduce the position dependence of absorber events. 
As the thermal conductivity of the gold strap, which is the thermalization layer, increases, the position dependence in the iron absorber becomes negligible because the heat is conducted to the TES before it escapes\cite{Mori+22}.

\begin{figure}[!htbp]
\centering
\vspace{-16mm}

\begin{minipage}{1\hsize} 
\centering
\includegraphics[angle=90, trim=0mm 0mm 0mm 0mm, width=0.8\linewidth, keepaspectratio]{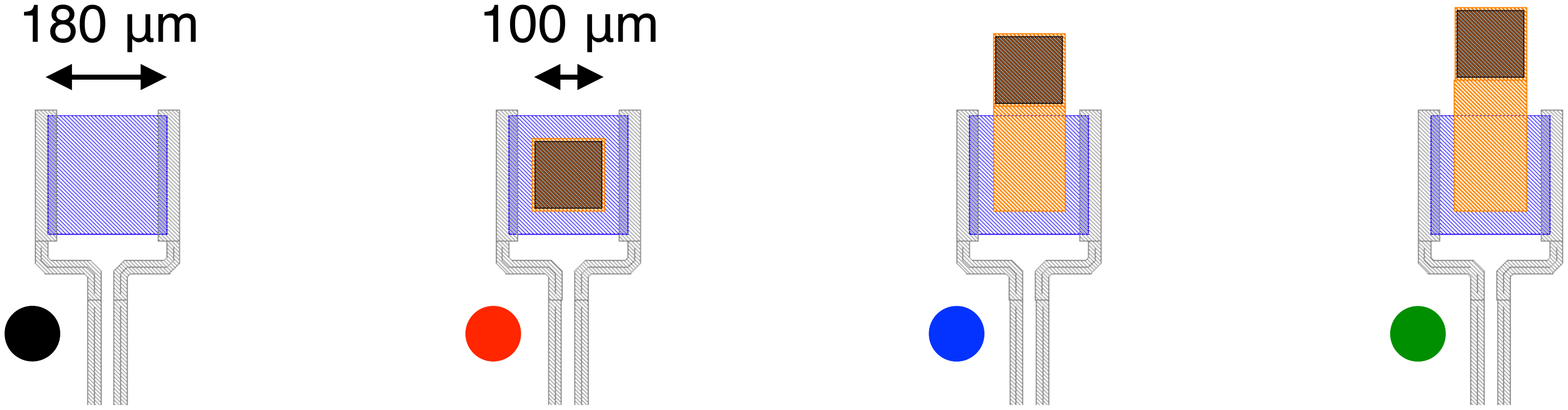}
\end{minipage}
\\
\vspace{-15mm}
\begin{minipage}{1\hsize} 
\centering
\includegraphics[trim=0mm 0mm 0mm 0mm, width=0.8\linewidth, keepaspectratio]{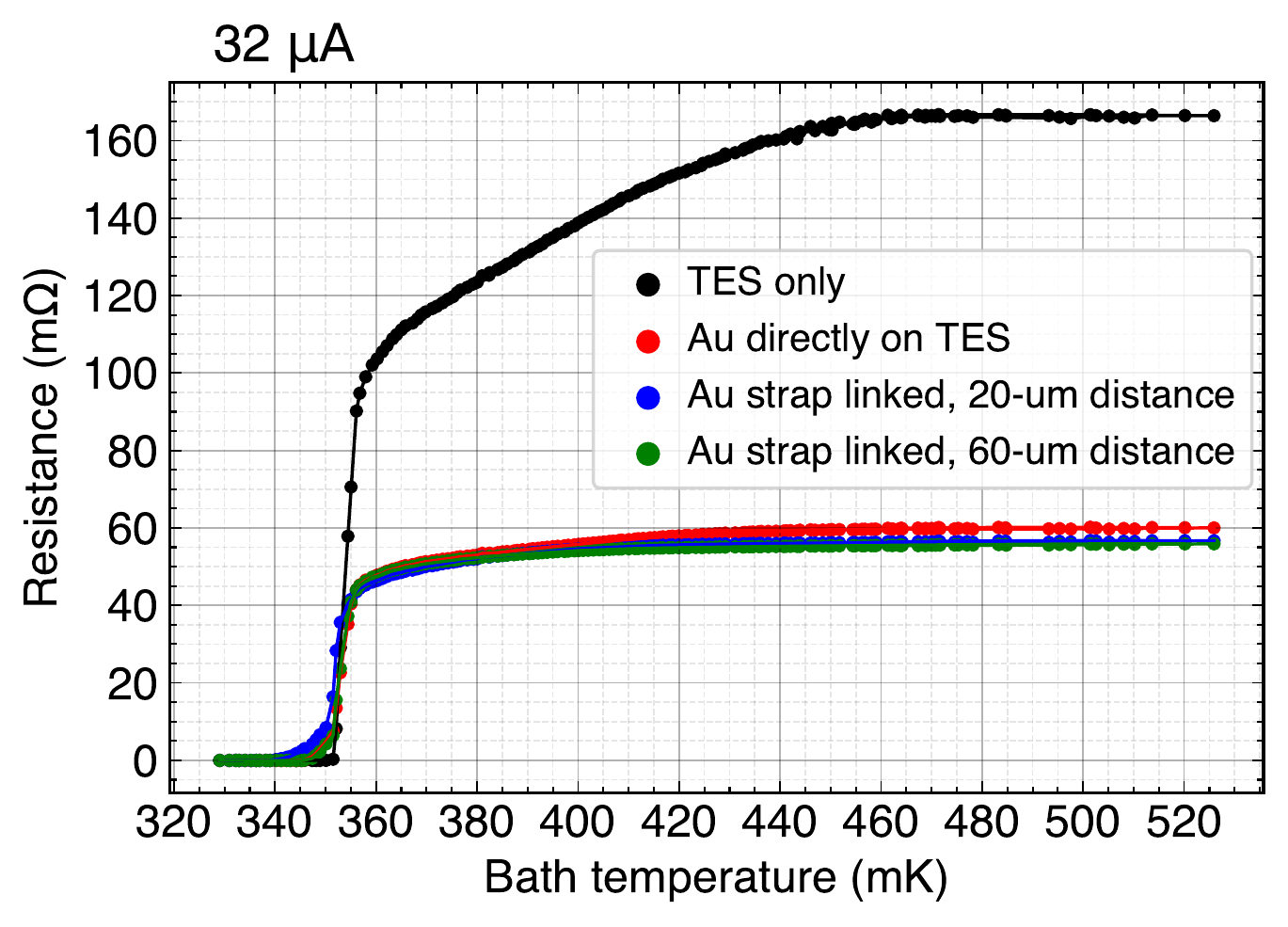}
\end{minipage}
\caption{(\textit{Top}) TES design with the thermal strap and iron absorber. The TES and iron absorber size  is $\num{180}$-$\si{\micro m}$ square and $\num{100}$-$\si{\micro m}$ square, respectively. The distance between the TES and absorber is $\SI{20}{\micro m}$ in the blue and $\SI{60}{\micro m}$ in the green, respectively.
 (\textit{Bottom}) RT curve of ``JAXA113 Ax1" with TES and thermal strap
structure. This was measured with increasing temperature.
Note that iron absorbers were NOT deposited in the elements using this RT measurement.}
\label{fig:RT}
\end{figure}

In order
to establish the conditions for the 
iron-absorber
deposition, 
the
TES was evaporated on a 3-inch substrate, 
the
niobium lead
($\sim \SI{250}{nm}$)
was sputtered, and the gold strap was plated. The 3-inch substrate was cut into two 35mm squares, and iron was plated on each. However, the gold solution used at this time had been open for a long time, and a high RRR was not able to be expected due to its lifetime. Nevertheless, this was okay since we would like to check the structure formation first. Therefore, the emphasis was placed on depositing gold with around $\num{2}$-$\si{\micro m}$ thickness without worrying about the current density in detail. Figure \ref{fig:RT} shows RT 
curves
with and without gold straps of 
$\SI{1.9}{\micro m}$ in thickness before the iron deposition. The transition characteristics were the same regardless of the strap size. 
The gold strap on the TES had low resistance, only about a few milliohms in size at the contact to TES. Therefore, where the strap was present, the current flew through the upper layer of the strap, not the lower layer of TES. As a result,
the normal resistance became smaller than that without the strap.

\section{Electroplating of Pure Iron for Absorbers}
The $^{57}$Fe used for the axion converter and absorber is very expensive, costing around 5,000 $\si{USD/g}$. 
Therefore, methods such as sputtering and vapor deposition, which coat the entire chamber, are unsuitable. 
Thus
we decided to use electroplating, which can deposit a film only on the necessary absorber pattern area, has high thermal conductivity, and has a low heat load during the process. However, since the deposition of pure iron, which is not an alloy, has not been done commercially or in basic research, we 
have been explored
the conditions for deposition in collaboration with
a
laboratory of Waseda University, which specializes in electrochemistry.

The drawbacks of pure iron deposition are that it is easily oxidized, it is difficult to control the film thickness on the order of several micrometers, and a low pH is required to dissolve the iron powder in the solution. If the pH is low, hydrogen generation tends to occur in the patterned area due to electric field concentration, inhibiting iron deposition. 
Futhermore,
it is desirable to be able to make a solution and deposit a film with as little iron powder as possible so as not to waste costly iron powder. For the initial stage of film formation, $^{56}$Fe was used instead of $^{57}$Fe, and iron (III) oxide (Fe$_2$O$_3$) powder on a $>$99.995\% trace metals basis from Sigma-Aldrich was used. The 
solution conditions
were referred to 
Inoue et al. 2002
\cite{Inoue+01}, and iron (III) oxide solutions of 0.01 mol/L and 0.02 mol/L were produced. We used 36\% hydrochloric acid (HCl) to dissolve the iron powder, boric acid (H$_3$BO$_3$) as a buffering agent to maintain a solution at a constant pH, and ammonium chloride (NH$_4$Cl) to assist ionic conduction. An aqueous solution of about $\SI{10}{mol/L}$ sodium hydroxide (NaOH) was added in a small amount at a time to prevent precipitation, and the pH was adjusted to about 1.5--2.0. Ultrapure water was added to bring the solvent volume to $\SI{140}{mL}$, and the air was degassed by bubbling with nitrogen gas for at least 20 minutes. The exact amounts of each solute and solvent are shown in tab. \ref{tab:Fe_solution_conditions}.

The counter electrode was connected to the Pt$/$Ti mesh, and the working electrode was connected to a patterned $\SI{35}{mm}$ square substrate and cut from a 3-inch substrate. Before deposition, O$_2$ ashing was performed to clean the substrate surface, followed by hydrochloric acid cleaning. The plated area of the pattern is $\SI{0.058}{cm^2}$, and an extra plated area of about $\SI{1}{cm^2}$ is provided around the perimeter of the substrate to reduce the concentration of the electric field on the pattern. The solution was heated to $\num{40}$--$\num{50}$$^\circ$C. The deposition was performed at a constant current value of $\SI{-40}{mA}$ with the stirring paddle ($\sim\SI{60}{rpm}$) in the vicinity of the substrate surface. The current density was approximately $\SI{-40}{mA/cm^2}$. The potential between the two poles was about $\SI{-2.2}{V}$.

Controlling the film thickness is an issue for the future. However, we have deposited a film of more than a few micrometers. The deposition conditions and the film thicknesses measured at several locations on the substrate are shown in tab. \ref{tab:Fe_ep_conditions}. It was difficult to measure the thickness of the Fe film immediately after deposition. Therefore, the film thickness was measured after the photoresist and seed layer was removed 
using the color 3D laser microscope VK-9710 (KEYENCE Corp.).
When the film thickness was about $\num{2}$--$\SI{3}{\micro m}$, there was a variation of at most $\pm$10\% from the average. In particular, when the film thickness was thicker than the protective photoresist ($\sim\SI{20}{\micro m}$), as in ``JAXA113 Ax1", not only did the film thickness tend to vary, but also a part of the surface tended to peel off during the process, resulting in a considerable significant variation.

\begin{figure}[!htbp]
\centering
\begin{minipage}{0.9\hsize} 
\begin{center}
\includegraphics[trim=0mm 0mm 0mm 0mm, width=0.9\linewidth, keepaspectratio]{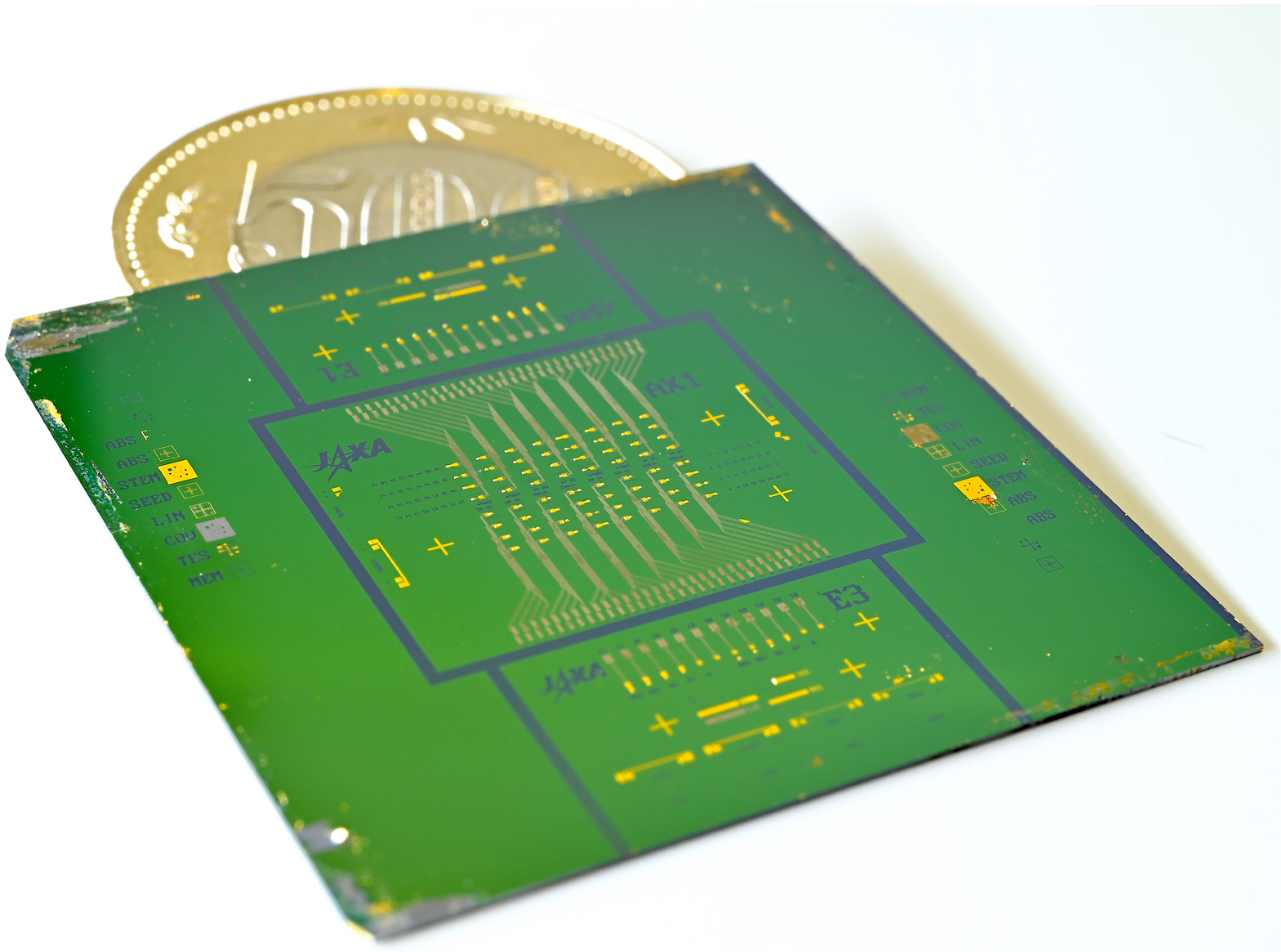}
\end{center}
\end{minipage}
\hspace{1mm}
\begin{minipage}{0.1\hsize} 
\begin{center}
\vspace{-25mm}
\hspace{-49mm}
\includegraphics[trim=0mm 0mm 0mm 0mm, width=3\linewidth, keepaspectratio]{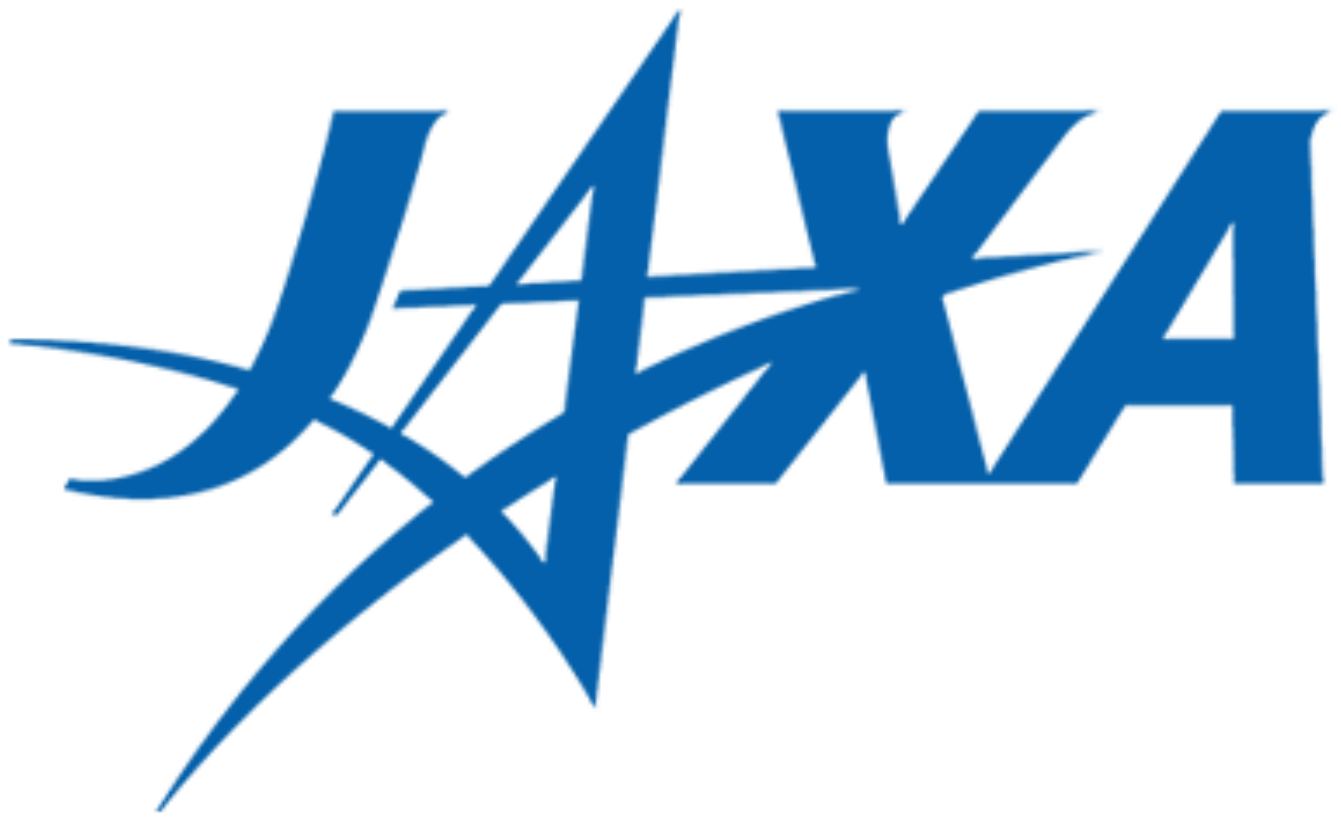}
\end{center}
\end{minipage}
\vspace{-10mm}
\caption{Our fabricated chip, ``JAXA105 Ax1," before forming the membrane structure with a Japanese 500-yen coin. The chip size is $\SI{35}{mm}$ square. The center section has 
a
64-pixel TES array with iron absorbers, and the up and down sections have 
a
12-pixel TES array to test the effect of the ferromagnetic iron field.}
\label{fig:JAXA105_Ax1}
\end{figure}

\begin{figure}[!htbp]
\centering
\begin{minipage}{0.9\hsize} 
\begin{center}
\includegraphics[angle=270, trim=0mm 0mm 0mm 0mm, width=0.9\linewidth, keepaspectratio]{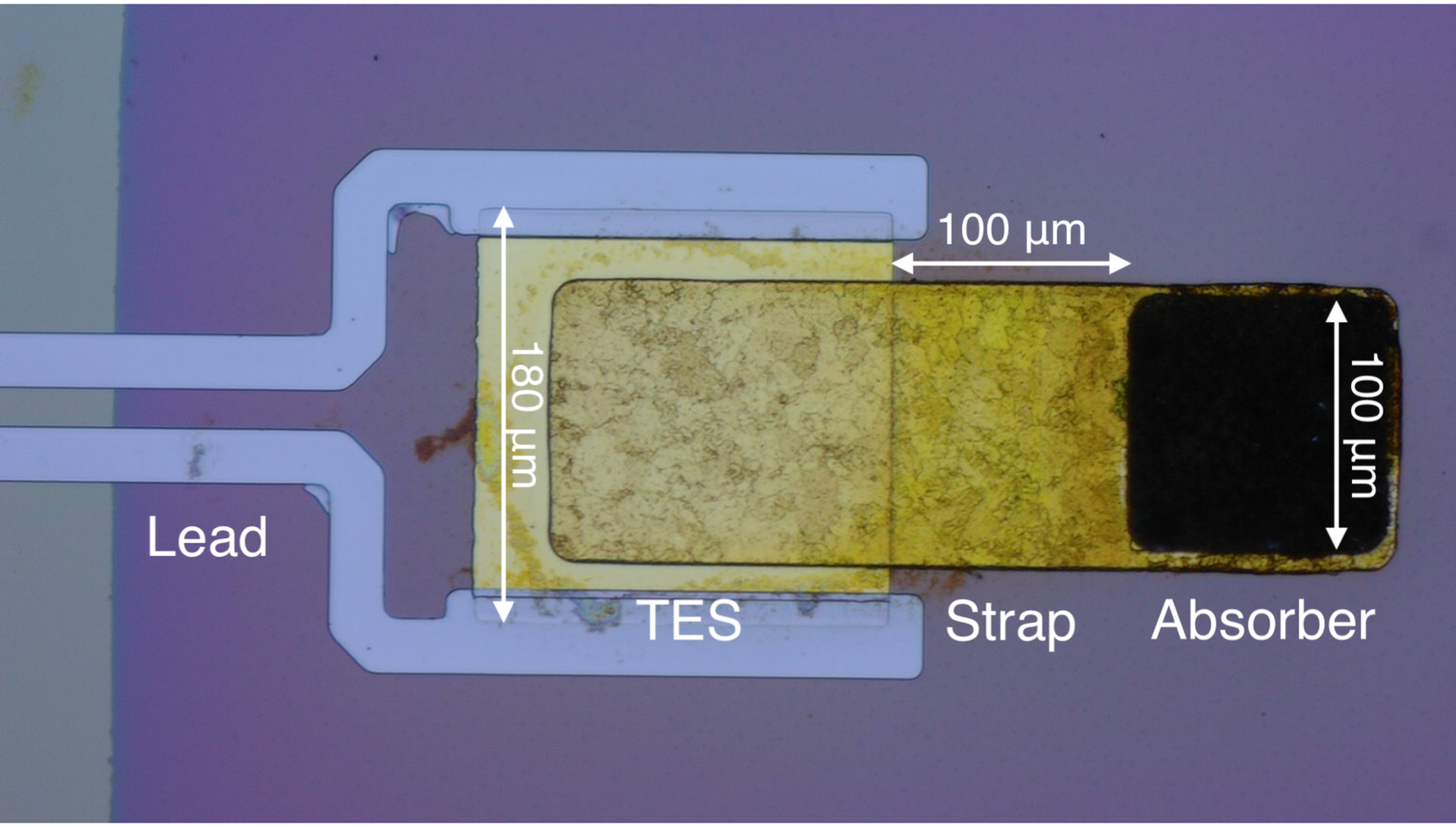}
\end{center}
\end{minipage}
\begin{minipage}{0.1\hsize} 
\begin{center}
\vspace{-24.2mm}
\hspace{-56.35mm}
\includegraphics[trim=0mm 0mm 0mm 0mm, width=2.3\linewidth, keepaspectratio]{jaxa_logo-2.pdf}
\end{center}
\end{minipage}
\vspace{-8mm}
\caption{An element of ``JAXA105 Ax1" with the iron absorber and membrane structure. The TES and iron absorber size was $\SI{180}{\micro m}$ square and $\SI{100}{\micro m}$ square, respectively. The distance between the TES and absorber is $\SI{100}{\micro m}$.}
\label{fig:JAXA105_E3_11}
\end{figure}

We successfully fabricated 
a
64-pixel TES calorimeter structure with iron absorbers for the first time. Figure \ref{fig:JAXA105_Ax1} shows 
the
64-pixel TES array. 
The TES and membrane structure
are
shown in fig. \ref{fig:JAXA105_E3_11}. We confirmed that the structure with only TES and thermal straps transitioned
between the normal and superconducting state
(fig. \ref{fig:RT}). However, a residual resistance of a few milliohms to a few ohms was observed 
in some pixels
after iron was deposited. The measured elements did not have a membrane structure yet. The 
residual
resistance could be due to the iron deposition process's effect or the iron's magnetism. Therefore, it is necessary to fabricate some more test elements and perform low-temperature measurements to clarify the cause. With a small residual resistance of 
less than a milliohm,
the effect of heat generation would be small, and there would be no problem operating as a calorimeter. We will also measure the degree of change in the TES sharpness due to the iron magnetism, which is more detailed than that of Konno et al. 2020 \cite{Konno+20}.

\begin{table}[!htbp]
\caption{Conditions for iron (III) oxide solution at concentrations of 0.01 mol/L and 0.02 mol/L}
\centering
\begin{tabular}{ccc}
\hline
Substance       & Concentration & Amount        \\ \hline
\multirow{2}{*}{Fe$_{2}$O$_{3}$} & 0.01 mol/L   & 0.2237 g      \\
                                 & 0.02 mol/L   & 0.4471 g      \\
HCl (aq)           & -             & 10 mL         \\
H$_{3}$BO$_{3}$ & 0.4 mol/L    & 3.4625 g      \\
NH$_{4}$Cl      & 2 mol/L     & 14.98 g       \\
NaOH (aq)       & -             & $\sim$ 10 mL  \\
Ultrapure water & -             & $\sim$ 120 mL \\ \hline
\end{tabular}
\label{tab:Fe_solution_conditions}
\end{table}

\begin{table}[!htbp]
\caption{Electroplating conditions and deposited iron thicknesses. Note that iron absorbers deposited on ``JAXA107 Ax1" without gold straps.}
\centering
\scalebox{0.9}[0.9]{
\begin{tabular}{cccccc}
\hline
ID                                                                                                    & \begin{tabular}[c]{@{}c@{}}JAXA105 \\ Ax1\end{tabular} & \begin{tabular}[c]{@{}c@{}}JAXA105 \\ Ax2\end{tabular} & \begin{tabular}[c]{@{}c@{}}JAXA107 \\ Ax1\end{tabular} & \begin{tabular}[c]{@{}c@{}}JAXA109 \\ Ax1\end{tabular} & \begin{tabular}[c]{@{}c@{}}JAXA113 \\ Ax1\end{tabular} \\  \hline
\vspace{+1mm}
\begin{tabular}[c]{@{}c@{}}
Iron \\ concentration \\ (mol/L)\end{tabular}                              & 0.01                                                   & 0.02                                                   & 0.01                                                   & 0.01                                                   & 0.02                                                   \\

\vspace{+1mm}
\begin{tabular}[c]{@{}c@{}}pH \end{tabular}                                               & 2.041                                                    & 1.613                                                     & 1.944                                                      & 1.855                                                     & 1.936                                                     \\

\vspace{+1mm}
\begin{tabular}[c]{@{}c@{}}Solution \\temperature \\($^\circ$C) \end{tabular}                                               & 47                                                     & 42                                                     & 49                                                      & 49                                                 & 41                                                     \\

\vspace{+1mm}
\begin{tabular}[c]{@{}c@{}}Current \\ (mA)\end{tabular}                                             & -40                                                    & -40                                                    & -40                                                    & -40                                                    & -40  
                                                  \\
\vspace{+1mm}
\begin{tabular}[c]{@{}c@{}}Time \\ (min)\end{tabular}                                               & 15                                                     & 15                                                     & 15                                                     & 15                                                     & 25                                                     \\
\multirow{6}{*}{\begin{tabular}[c]{@{}c@{}}Measured\\  thickness \\ ($\si{\micro m}$)\end{tabular}} & 3.157                                                  & 2.216                                                  & 3.133                                                  & 2.580                                                  & 9.532                                                  \\
                                                                                                    & 3.207                                                  & 2.226                                                  & 3.725                                                  & 2.706                                                  & 20.470                                                 \\
                                                                                                    & 3.484                                                  & 2.246                                                  & \multicolumn{1}{l}{}                                   & 3.009                                                  & 22.426                                                 \\
                                                                                                    & 3.630                                                  & 2.340                                                  & \multicolumn{1}{l}{}                                   & 3.039                                                  & 57.798                                                 \\
                                                                                                    & 3.649                                                  & 2.417                                                  & \multicolumn{1}{l}{}                                   & \multicolumn{1}{l}{}                                   & 62.365                                                 \\
                                                                                                    & \multicolumn{1}{l}{}                                   & \multicolumn{1}{l}{}                                   & \multicolumn{1}{l}{}                                   & \multicolumn{1}{l}{}                                   & 62.491                                                 \\
\vspace{+1mm}
\begin{tabular}[c]{@{}c@{}}Averaged \\ thickness\\ ($\si{\micro m}$)\end{tabular}                   & 3.425                                                  & 2.289                                                  & 3.429                                                  & 2.834                                                  & 39.180                                                 \\
\vspace{+1mm}
\begin{tabular}[c]{@{}c@{}}Sample \\ standard \\ deviation\\ ($\si{\micro m}$)\end{tabular}         & 0.232                                                  & 0.087                                                  & 0.419                                                  & 0.226                                                  & 24.238                                                 \\ \hline
\end{tabular}
}
\label{tab:Fe_ep_conditions}
\end{table}

\section{Conclusion and Future Work}
We introduced the vacuum evaporator and found the 
linear
trend between the ratio of Au to Ti thickness and transition temperature. We electroplated gold straps with high thermal conductivity 
(RRR$>$23)
and absorbers of more than a few micrometers
in thickness
for absorbers using iron (III) oxide powder. For the first time, the fabrication of the 64-pixel TES array with iron absorbers was achieved. However, the residual resistance appeared in the elements with the iron absorber. Therefore, a further test device will be fabricated, and the influence of iron magnetism on TES will be studied in detail.

\vfill


\begin{thebibliography}{1}
\bibliographystyle{IEEEtran}


\bibitem{PecceiandQuinn77} 
R. D. Peccei and H. R. Quinn, ``CP Conservation in the Presence of Pseudoparticles," {\it Phys. Rev. Lett.}, \textbf{38}, p. 1440, 1977, \url{https://doi.org/10.1103/PhysRevLett.38.1440};\\
R. D. Peccei and H. R. Quinn, ``Constraints imposed by CP conservation in the presence of pseudoparticles," {\it Phys. Rev. D}, \textbf{16}, p. 1791, 1977, \url{https://doi.org/10.1103/PhysRevD.16.1791}

\bibitem{WeinbergandWilczek78} 
S. Weinberg, ``A New Light Boson?," {\it Phys. Rev. Lett.}, 1978, \textbf{40}, p. 223, \url{https://doi.org/10.1103/PhysRevLett.40.223};\\
F. Wilczek, ``Problem of Strong P and T Invariance in the Presence of Instantons," {\it Phys. Rev. Lett.}, \textbf{40}, p. 279, 1978, \url{https://doi.org/10.1103/PhysRevLett.40.279}


\bibitem{Turck+93} 
S. Turck-Chi$\grave{\rm e}$ze, W. D$\ddot{\rm a}$ppen, E. Fossat, J. Provost, E. Schatzman, and D. Vignaud, ``The solar interior," {\it Phys. Rep.}, \textbf{230}, p. 57, 1993, \url{https://doi.org/10.1016/0370-1573(93)90020-E}

\bibitem{Yagi+23} 
Y. Yagi, R. Konno, T. Hayashi, K. Tanaka, N. Y. Yamasaki, K. Mitsuda, R. Sato, M. Saito, T. Homma, Y. Nishida, S. Mori, N. Iyomoto, T. Hara, ``Performance of TES X-Ray Microcalorimeters Designed for 14.4-keV Solar Axion Search," {\it J. Low Temp. Phys.}, published online, 2023, \url{https://doi.org/10.1007/s10909-023-02942-w}


\bibitem{Moriyama95} 
S. Moriyama, ``Proposal to Search for a Monochromatic Component of Solar Axions Using 57 Fe," {\it Phys. Rev. Lett.}, \textbf{75}, p. 3222, 1995, \url{https://doi.org/10.1103/PhysRevLett.75.3222}

\bibitem{Krcmar+98} 
M. Kr$\check{\rm c}$mar, Z. Kre$\check{\rm c}$ak, M. Stip$\check{\rm c}$evi$\acute{\rm c}$, A. Ljubi$\check{\rm c}$i$\acute{\rm c}$, and D. A. Bradley, ``Search for Solar Axions Using 57Fe," {\it Phys. Lett. B}, \textbf{442}, p. 38, 1998, \url{https://doi.org/10.1016/S0370-2693(98)01231-3}

\bibitem{Derbin+07} 
A. V. Derbin, A. I. Egorov, I. A. Mitropol’sky, V. N. Muratova, N. V. Bazlov, S. V. Bakhlanov, D. A. Semenov, and E. V. Unzhakov, ``Search for resonant absorption of solar axions emitted in an M1 transition in 57Fe nuclei," {\it JETP Lett.}, \textbf{85}, p. 12, 2007, \url{https://doi.org/10.1134/S0021364007010031}

\bibitem{Namba07} 
T. Namba, {\it Phys. Lett. B}, ``Results of a search for monochromatic solar axions using 57 Fe," \textbf{645}, p. 398, 2007, \url{https://doi.org/10.1016/j.physletb.2007.01.005}

\bibitem{Derbin+09} 
A. V. Derbin, A. I. Egorov, I. A. Mitropolsky, V. N. Muratova, D. A. Semenov, and E. V. Unzhakov, ``Search for resonant absorption of solar axions emitted in M1 transition in 57Fe nuclei," {\it Eur. Phys. J. C}, \textbf{62}, p. 755, 2009, \url{https://doi.org/10.1140/epjc/s10052-009-1095-y}

\bibitem{Derbin+11} 
A. V. Derbin, V. N. Muratova, D. A. Semenov, and E. V. Unzhakov, ``New limit on the mass of 14.4-keV solar axions emitted in an M1 transition in 57Fe nuclei," {\it Phys. Atom. Nucl.}, \textbf{74}, p. 596, 2011, \url{https://doi.org/10.1134/S1063778811040041}

\bibitem{K79SVZ80} 
J. E. Kim, ``Weak-Interaction Singlet and Strong CP Invariance," {\it Phys. Rev. Lett.} \textbf{43}, p. 103, 1979, \url{https://doi.org/10.1103/PhysRevLett.43.103};\\
M. A. Shifman, A. I. Vainshtein, and V. I. Zakharov, ``Can confinement ensure natural CP invariance of strong interactions?," {\it Nucl. Phys. B}, \textbf{166}, p. 493, 1980, \url{https://doi.org/10.1016/0550-3213(80)90209-6}

\bibitem{Gavrilyuk15} 
Y. M. Gavrilyuk, A. N. Gangapshev, A. V. Derbin, I. S. Drachnev, V. V. Kazalov, V. V. Kobychev, V. V. Kuz’minov, V. N. Muratova, S. I. Panasenko, S. S. Ratkevich, D. A. Semenov, D. A. Tekueva, E. V. Unzhakov, and S. P. Yakimenko, ``New experiment on search for the resonance absorption of solar axion emitted in the M1 transition of 83Kr nuclei," {\it JETP Lett.}, \textbf{101}, p. 664, 2015, \url{https://doi.org/10.1134/S0021364015100069}

\bibitem{Gavrilyuk18} 
Y. M. Gavrilyuk, A. N. Gangapshev, A. V. Derbin, I. S. Drachnev, V. V. Kazalov, V. V. Kobychev, V. V. Kuzminov, V. N. Muratova, S. I. Panasenko, S. S. Ratkevich, D. A. Tekueva, E. V. Unzhakov, and S. P. Yakimenko, ``New Constraints on the Axion-Photon Coupling Constant for Solar Axions," {\it JETP Lett.}, \textbf{107}, p. 589, 2018, \url{https://doi.org/10.1134/S0021364018100090}






\end{thebibliography}
\end{document}